\documentclass{elsarticle}
 \usepackage{amssymb}
 \usepackage{graphicx}

\begin{document}

\title{Adiabatic dynamics of one-dimensional classical Hamiltonian dissipative systems}


\author[IRE]{G.M.~Pritula}
\ead{pritula.galina@gmail.com}
\author[Univer]{ E.V.~Petrenko}
\ead{petrenkoyevgen@yandex.ua}
\author[IRE]{O.V.~Usatenko}
\address[IRE]{A.Ya. Usikov Institute for Radiophysics and
Electronics, Ukrainian Academy of Science, 12 Proskura Street, 61085
Kharkov, Ukraine}
\address[Univer]{V.N. Karazin Kharkiv National University, 4
Freedom Square, 61077, Kharkiv, Ukraine}

\begin{abstract}

We give an example of a simple mechanical system described by the
generalized harmonic oscillator equation, which is a basic model in
discussion of the adiabatic dynamics and geometric phase. This
system is a linearized plane pendulum with the slowly varying mass
and length of string and the suspension point moving at a slowly
varying speed, the simplest system with broken $T$-invariance. The paradoxical
character of the presented results is that the same Hamiltonian
system, the generalized harmonic oscillator in our case, is
canonically equivalent to two different systems: the usual plane
mathematical pendulum and the damped harmonic oscillator. This once
again supports the important mathematical conclusion, not widely
accepted in physical community, of no difference between the
dissipative and Hamiltonian 1D systems, which stems from the Sonin
theorem that any Newtonian second order differential equation with a
friction of general nature may be presented in the form of the
Lagrange equation.

\end{abstract}

\begin{keyword}
adiabatic dynamics\sep geometric phase\sep Lagrangian and Hamiltonian mechanics\sep dissipative systems
\PACS
01.55.+b \sep 03.65.Vf \sep 45.20.Jj \sep 45.30.+s \sep 45.50.Dd
\end{keyword}

 \maketitle

\section{Introduction\label{Intro}}
Dynamics of our world is governed and described by differential
equations. Realization of this startling fact was evaluated by
Newton as the most important discovery of his life. However,
explicit analytical solutions of differential equations are the
exception rather than the rule. This makes scientists develop
special and approximate methods for the analysis of differential
equations because every new step in understanding the properties of
their solutions gives a further insight into a physical theory described by
corresponding equations. Thus, for example, the discovery of
adiabatic invariants of the second order differential equation with
slowly varying parameters was an important step in the development
of quantum theory. The existence of one more remarkable property of
this equation, the so-called geometric phase, was noticed only 80
years later. Historical aspects of the development of ideas related
to the understanding of the properties of solutions of differential
equations with slowly varying parameters as well as their
theoretical, experimental and applied aspects one can find in many
reviews and books (see, for example,
\cite{Wilczek,Vinitskii,Klyshko,Malykin,Chruscinski}).

The quantity considered in the present paper, the geometric phase,
is also known as the topological or nonholonomic phase and often
associated with the names of its pioneers: Rytov, Vladimirskii,
Pancharatnam, Berry, Hannay, less frequently with Ishlinskii (who
gave the explanation of systematic gyroscope bias error after a
long voyage), and others. In our work we consider this concept at
the classical, non-quantum level and in what follows call it the
geometric or Hannay phase.  The geometric phase can occur both in
quantum \cite{Berry84} and in classical \cite{Hannay,Berry85}
systems. This is not astonishing in view of the actually identical
second order differential equations which are the time-independent
Schr$\ddot{o}$dinger equation and the Newton (or Hamilton)
equation for the harmonic oscillator with a linear restoring force.
The analogy between quantum and classical phenomena is clearly seen, for
example, when one compares the classical phenomenon of parametric
resonance and the band character of the spectrum of quantum particle
in a stationary periodic field: both of the phenomena are described
by the Hill equation. This analogy was also repeatedly used  in the
study and comparison of the adiabatic dynamics of classical systems
and the WKB approximation of quantum mechanics \cite{Dykhne60}. In
mathematical terms, the geometric phase is a correction to the
dynamical phase for the harmonic solution of a linear differential
equation with the broken time-reversal invariance or, in other
words, for the solution which describes the vibrational mode of
motion of dynamical systems \cite{Fedoruk,Bliokh} with slowly
varying parameters.

In the present work we give an elementary example of mechanical
system illustrating the physical meaning of Hamiltonian
(\ref{Hamiltonian1}) and, in this way, the possible range of
applicability of Hannay's \cite{Hannay} results.  This mechanical
system is a plane mathematical pendulum with the slowly varying mass
and string length, and with the suspension point moving at a slowly
varying speed. The fact of canonical equivalence between the
considered pendulum and a damped harmonic oscillator is surprising
from the physical point of view and trivial, at the same time, from
mathematical point. We discuss this duality at Section
\ref{Discuss}. A complex form of GHO Hamilton function is presented
in Appendix.

\section{Generalized harmonic oscillator \label{GHO}}

The simplest second order equation which can be a
demonstrative example of the existence of geometric phase in
classical adiabatic dynamics
\cite{Hannay,Berry85} is the equation of motion of the generalized
harmonic oscillator (GHO). The importance of this example is confirmed
by the fact that scientists after Hannay \cite{Hannay} often
returned to this equation
\cite{Berry85,Bliokh,Usatenko,Razavy,Smith,Kobe} using different
methods for its analysis. The Hamiltonian of the
GHO is given by
\begin{equation}
    H=
        \frac{1}{2}\left(
        \alpha Q^2+2\beta QP+\gamma P^2
        \right),
\label{Hamiltonian1}
\end{equation}
where $Q$ and $P$ are the canonically conjugate coordinate and
momentum; $\alpha$, $\beta$, and $\gamma$  are the parameters of the
generalized oscillator. When the parameters $\alpha$, $\beta$ and
$\gamma$ are constant, the energy of the system is a constant of
motion. For the
values of $\alpha$,
$\beta$ and
$\gamma$ satisfying the inequality $\alpha\gamma>{\beta}^2$,
solutions of the Hamilton equations take the form:
\begin{equation}
    Q=r \cos\Theta,\;
    P=-\frac{r}{\gamma}\left(
        \beta \cos\Theta+\omega\sin\Theta
        \right),\;
        \Theta=\omega t,\;
    \omega=\sqrt{\alpha\gamma-{\beta}^2}.
    \label{Solutions1}
\end{equation}


%


If the parameters change slowly, see Eq. (\ref{Epsilon}), the motion
of the oscillator can be approximately
regarded as the periodic one of the form (\ref{Solutions1}) with the
slowly varying amplitude $r$ and phase $\Theta$ both of which should
be determined. In this case, the energy of the system is not
conserved, but there is a new approximate conserved quantity, the
adiabatic invariant,
\begin{equation}
        I=\frac{\omega r^2}{2\gamma}=\frac{E}{\omega},
        \label{AdInvar}
\end{equation}
which remains constant with (non-analytic) exponential accuracy
$\Delta I\sim \exp(-1/t_0 \epsilon)$ \cite{Slutskin64},
where $t_0$ is some constant
determined  by analytical properties of varying parameters.
The phase of the oscillator is an `almost linear' function of time. It
was shown in the works by Hannay \cite{Hannay} and Berry \cite{Berry85}
that the phase of the oscillator can be represented as the sum of two
quantities, $\Theta=\Theta_d+\Theta_g$, where the dynamic $\Theta_d$
and the geometric $\Theta_g$ phases are:
\begin{equation}
    \Theta_d=\int^t\!\!{\omega dt},\quad
    \Theta_g=
    \frac{1}{2}\int^t{\frac{\beta}{\omega}}
    \left(
        \frac{\dot{\gamma}}{\gamma}-\frac{\dot{\beta}}{\beta}
    \right) dt=
    \frac{1}{2}\int_{\Gamma}
        {\frac{\beta}{\omega}}
        \left(
        \frac{d\gamma}{\gamma}-\frac{d\beta}{\beta}
        \right).
    \label{D&GPhases}
\end{equation}

The independence of the geometric phase of time follows from the last equation
of (\ref{D&GPhases}), provided the adiabatic condition holds, and
explains the name of this phase. The dependence of the phase on the
path $\Gamma$ of integration is associated with the concept of
anholonomy. Reversing the direction of integration  along the
contour $\Gamma$ changes  the sign of the geometric phase, and when
the parameter $\beta$, which violates the time-reversal invariance
of the Hamiltonian (\ref{Hamiltonian1}), becomes zero, the geometric
phase vanishes as well. If the line of variation of the parameters
$\alpha$, $\beta$ and $\gamma$ is the closed curve $\Gamma$, then
the integral corresponding to the geometric phase can be converted
by Stokes' theorem to the surface integral which is also independent
of the time of parameter change:
\begin{equation}
        \Theta_g=\int\!\!\!\int_{S(\Gamma)}{\frac{1}{4\omega^3}}\left(
        \gamma dS_{\alpha\beta}+\alpha dS_{\beta\gamma}+\beta dS_{\alpha\gamma}
        \right),
    \label{GPhase}
\end{equation}
where $dS_{\alpha\beta}=d\alpha\wedge d\beta$ and similar
expressions are projections of oriented surface elements on relevant
directions. This is the essence of Berry and Hannay's results in the
application to the classical mechanics. The result (\ref{D&GPhases})
can be obtained \cite{Usatenko} by averaging over the `fast'
variable $\Theta$ without appealing
to the action-angle variables, as it was originally made in Hannay's
work \cite{Hannay} . In more details this procedure is as follows: it
is necessary to substitute the expressions (\ref{Solutions1}) into
the Hamiltonian equations of motion,
\begin{equation}
        \dot P=-\frac{\partial H}{\partial Q},\quad
        \dot Q= \frac{\partial H}{\partial P},
        \label{HamEqs}
\end{equation}
bearing in mind that the parameters $\alpha$, $\beta$ and $\gamma$
are functions of time. Solving the obtained equations with
respect to $\dot r$ and $\dot \Theta$ and averaging them over the
period of motion, one arrives at simple differential
 equations for $\alpha$, $\beta$ and $\gamma$ that have the solutions given by
the expressions (\ref{AdInvar}) and (\ref{D&GPhases}). Another
alternative method of obtaining the results (\ref{AdInvar}) and
(\ref{D&GPhases}) one can find in Appendix B.

Note, that the geometric phase $\Theta_g$ stems from the non-potential
(vortex) nature
of the differential form
$\frac{\beta}{\omega}\left(\frac{d\gamma}{\gamma}-\frac{d\beta}{\beta}\right)$,
since in general case
$\frac{\partial}{\partial\gamma}\left(-\frac{1}{\omega}\right)
\neq\frac{\partial}{\partial\beta}\left(\frac{\beta}{\gamma\omega}\right)$;
in the theory of differential forms such forms are called inexact. The
Hannay phase cannot be calculated only on the basis of initial and
final states of the oscillator and depends on the path connecting
the start $(\alpha_s,\beta_s,\gamma_s)$ and end
$(\alpha_e,\beta_e,\gamma_e)$ points-states of the system in the
parameter space $(\alpha,\beta,\gamma)$.  For the existence of the geometric
phase (see  Eq. (\ref{D&GPhases})), the most significant
factor is the lack of $T$-invariance of Hamiltonian
(\ref{Hamiltonian1}).

In spite of the simplicity, this result had a great influence on
the subsequent development of the theory of dynamical systems and
found numerous applications \cite{Wilczek,Vinitskii,Klyshko,Malykin,
Chruscinski}. However, until now the question,
which systems can be described by the Hamiltonian of the GHO
is still open. In the work \cite{Usatenko} it was
shown that the Hamiltonian (\ref{Hamiltonian1}) is canonically
equivalent to the Hamiltonian of the equation of damped harmonic
oscillator. The result, on the one hand, is a bit surprising but, on
the other hand, leaves a feeling of dissatisfaction. In particular,
the existence of other simple counterparts of the GHO
among well-known systems of mechanical or other
origin seems to be natural.



\section{Plane mathematical pendulum \label{MathPend}}

\begin{figure}
\hspace{25mm}
\includegraphics[height=6cm]{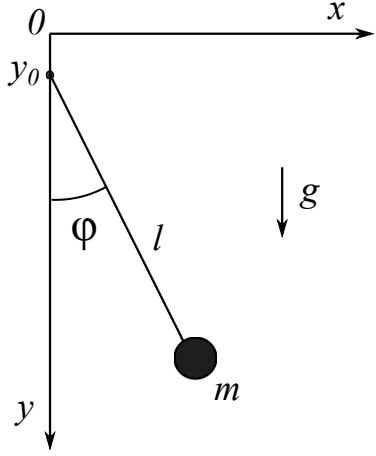}
\caption{A pendulum oscillating in the $XOY$ plane with the
suspension
point moving along the axis $OY$.}%
\label{Fig2}%
\end{figure}

Let us consider the motion of simple plane linearized mathematical
pendulum with the suspension point moving with a small acceleration
along the vertical axis $OY$ of the oscillation plane $XOY$, see
Fig.\ref{Fig2}. The speed $v$ of the suspension point as well as two
other pendulum parameters -- the mass $m$ and  the length $l$
of the string  -- are supposed to be slowly changing functions of time
with the characteristic scale $\epsilon^{-1}$ much greater
than the period $T$ of harmonic oscillations of the pendulum:
\begin{equation}
    \epsilon T <<1.
\label{Epsilon}
\end{equation}

For small oscillations, the coordinates of the point $m$ are:
\begin{equation}
    x=l \sin\phi\approx l\phi,\quad
    y=y_0 +l \cos\phi\approx y_0 +l(1-{\phi}^2/2).
\label{Coordinates}
\end{equation}
Here $y_0$  is the coordinate of the suspension point of the
pendulum and $\phi$  is the deflection angle of the pendulum.
The Lagrangian of the pendulum up to the terms of
the first order in $\epsilon
T<<1$ can be written as
\begin{equation}
    L=\frac{ml^2}{2}{\dot{\phi}}^2-mlv\phi\dot{\phi}-\frac{ml}{2}
    \left(
    g+v\frac{\dot l}{l}
    \right)
    {\phi}^2,
\label{Lagrangian}
\end{equation}
where $v=\dot{y_0}$. Using the standard
Legendre transformation and introducing the generalized momentum
$p=\partial L/\partial\dot{\phi}$ instead of generalized velocity $\dot{\phi}$,
we find the Hamilton function $H=p\dot{\phi}-L$,
\begin{equation}
        H=\frac{1}{2ml^2}p^2+
        \frac{v}{l}p\phi+
        \frac{ml}{2}
        \left(
        g+\frac{v^2}{l}+v\frac{\dot l}{l}
        \right)
        \phi^2.
\label{Hamiltonian2}
\end{equation}

Thus, we directly arrive at the Hamilton (\ref{Hamiltonian1}) of the
GHO, where the parameters of the
Hamiltonian are given by:
\begin{equation}
        \alpha=m(gl+v^2+v\dot{l}),\
        \beta=v/l,\
        \gamma=1/ml^2.
        \label{Param1}
\end{equation}

The second term in (\ref{Hamiltonian2}) arises due to the transition to
the moving frame of reference
and the lack of invariance under the spatial transformation $y\rightarrow -y$.
(Such term does not appear when the suspension point of the pendulum moves only along the horizontal axis.)
The renormalization of gravity $g$ in
the third term of (\ref{Hamiltonian2}) is related to the action of
forces emerging in the non-inertial frame of reference.

Substituting (\ref{Param1}) into (\ref{AdInvar}) and
(\ref{GPhase}), we immediately obtain the adiabatic invariant,
\begin{equation}
        I=
                \frac{1}{2}mg^{1/2}l^{3/2}r^2,
        \label{Action}
\end{equation}
and both (the dynamical and geometric) phases of the pendulum:
\begin{equation}
    \Theta_d=\int^t\!\!{\sqrt{\frac{g}{l}} dt},\quad
    \Theta_g=
    -\int_{\Gamma}
        \frac{v}{2\sqrt{gl}}\left(\frac{dm}{m}+\frac{dv}{v}\right).
    \label{D&GPhasesPend}
\end{equation}

We would like to note here that the parameters $\alpha, \beta$ and $\gamma$
of the GHO play different role  in determining the geometric phase.
Indeed, the expression for the
geometric phase in (\ref{D&GPhases}) does not contain the derivative
of $\alpha$  with respect to $t$. By this reason we should take into
account the small parameter $\dot l$ in $\alpha$ and in Eqs.
(\ref{Lagrangian}) and (\ref{Hamiltonian2}), and neglect terms with
such derivatives in $ \beta$ and $\gamma$, if there are any. In
view of the equality $\omega=\sqrt{\alpha\gamma-{\beta}^2}$, the
expression for the dynamical phase also contains the parameter
$\alpha$ and, in this way, the term with $\dot{l}$, which should
obviously be
included into the geometric phase $\Theta_g$.

As can be seen from Eqs. (\ref{Hamiltonian2}) and
(\ref{D&GPhasesPend}), the variation of parameter $v$ is crucial for
the occurrence of the geometric phase. If the suspension point of
pendulum moves with a constant speed, the geometric phase $\Theta
_g=0$. The second term in (\ref{Hamiltonian2}) is responsible for
the absence of time reversal symmetry of the system. Variations of
only $l$ and $m$ cannot give rise to the geometric phase. The
variation of $l$ in the second term of Eq. (\ref{Hamiltonian2}) is
not enough for engendering the geometric phase because the terms
proportional to the time derivative of $l$ in the second and third
terms of (\ref{Hamiltonian2}) cancel each other.

Thus, we have shown the canonical equivalence of the simple
linearized mathematical pendulum and the GHO with adiabatically
varying parameters.

\section{Damped harmonic oscillator \label{Damped}}

In the work \cite{Usatenko} it was shown that the Hamiltonian
(\ref{Hamiltonian1}) is canonically equivalent to the Hamiltonian of the
equation for the dissipative harmonic oscillator:
\begin{equation}
        \frac{1}{M} \frac{d}{dt}(M\dot{q})+2\lambda\dot{q}+{\Omega}^2 q=0.
                \label{EqDampedHO}
\end{equation}

Indeed, it is easy to check that (\ref{EqDampedHO}) follows from the
Euler-Lagrange equation along with the generalized Caldirola-Kanai
Lagrangian  \cite{Usatenko,Caldirola,Kanai},
\begin{equation}
L(q,\dot{q},t)= \frac{M}{2} \exp{
    \left[2\int^t \lambda(s)ds
    \right]}
    ({\dot{q}}^2-{\Omega}^2 q^2).
\label{CaldirolaKanaiLag}
\end{equation}
The Lagrangian (\ref{CaldirolaKanaiLag}) makes it possible to
calculate the Hamiltonian corresponding to Eq.
(\ref{EqDampedHO}):
\begin{equation}
H(p,q)= \frac{p^2}{2M} \exp{
    \left[-2\int^t \lambda(s)ds
    \right]}+
    \frac{M{\Omega}^2 q^2}{2}\exp{
    \left[2\int^t \lambda(s)ds
    \right]}.
        \label{CaldirolaKanaiHam}
\end{equation}

The equations known from the theory of canonical transformations,
\begin{equation}
    p=\frac{\partial F}{\partial q},\ \
    Q=\frac{\partial F}{\partial P},\ \
    H(P,Q)=H(p,q)+\frac{\partial F}{\partial t},
        \label{CanonicTrans}
\end{equation}
establish the canonical equivalence of two different ways of
describing the same Hamiltonian system. The generating function
\begin{equation}
F(q,P,t)= qP\exp{
    \left[\int^t \lambda(s)ds
    \right]},
\label{GeneratingF}
\end{equation}
and the expression (\ref{CanonicTrans}), after the identification of
the parameters
\begin{equation}
    \alpha=M{\Omega}^2,\ \beta=\lambda,\ \gamma=M^{-1},
\label{CorrespondenceDO&GHO}
\end{equation}
allow us to verify the equivalence of (\ref{Hamiltonian1})
and (\ref{CaldirolaKanaiHam}) by direct calculation.

Thus, the equivalence of the plane pendulum to the GHO, on the one
hand, and the GHO to the damped harmonic oscillator (DHO), on the
other hand, yield the equivalence of the plane pendulum to the DHO.
The last conclusion can be verified directly by the substitution
\begin{equation}
   q=\phi\exp{
    \left[-\int^t \lambda(s)ds
    \right]};\,\, M=ml^2,\ \lambda=v/l,\ {\Omega}^2=(gl+v^2+v\dot{l})l^{-2},
\label{CorrespondenceDO&Pendulum2}
\end{equation}
into the Lagrangian (\ref{CaldirolaKanaiLag}). As a result we obtain
expression (\ref{Lagrangian}).

To obtain Caldirola-Kanai Lagrangian (\ref{CaldirolaKanaiLag}) of
the damped oscillator we have to put
\begin{eqnarray}
        \phi=q\exp{
    \left[\int^t \lambda(s)ds
    \right]}
\label{CorrespondenceDO&Pendulum1}
\end{eqnarray}
into (\ref{Lagrangian}) and find $\lambda$ by eliminating the
term with $q\dot{q}$. This results in $\lambda=v/l$.

Thus, we have shown that the planar pendulum (\ref{Lagrangian}) is
canonically equivalent to the dissipative DHO (\ref{EqDampedHO}),
(\ref{CaldirolaKanaiLag}). The equivalence holds for the systems
with slowly varying parameters (not only for the constant ones).
Another complex form of the GHO Hamiltonian, which is also
canonically equivalent to the original GHO Hamiltonian
(\ref{Hamiltonian1}) is presented in Appendix A.

\section{ Discussion \label{Discuss}}

Let us expose two different points of view on the subject under
discussion. Here we call them "physical" and
"mathematical".

The physicists  believe that dissipative forces are outside of the
scope of applicability of variational principles of analytical
mechanics \cite{Landau,Lanczos}. The general point of view on
applicability of this principle may be represented as follows (see,
e.g., \cite{Tarasov94,Tarasov08}). Newtonian 'vector' mechanics
describes the motion of mechanical systems under the action of
forces applied to them. Newton's approach does not limit the nature
of the forces, which are usually divided into potential and
dissipative. The Lagrange-Hamilton variational mechanics describes
the motion of mechanical systems under the action of only potential
forces \cite{Lanczos,Goldstein}. From this point the existence of
Lagrangian for dissipative system and its equivalence to
non-dissipative one is something extraordinary. Nevertheless the
Caldirola-Kanai Lagrangian describing the dissipative oscillator is
not an exceptional example of `exotic'  systems (with the equation
of motion containing the time derivatives of generalized
coordinates). The Caldirola-Kanai Lagrangian also describes systems
with an arbitrary potential $U(q)$ instead of $M{\Omega}^2q^2/2$ in
(\ref{CaldirolaKanaiLag}).

Other example of the equation of motion and the Lagrange function
for the oscillator with quadratic dependence of `friction' on velocity
reads \cite{Smith}:
\begin{equation}
    \ddot{q}+b{\dot{q}}^2 + \omega^2 q=0,
    \label{FiPoitSquarEq}
\end{equation}
\begin{equation}
    L=\frac{m}{2}\left[\left(\dot{q}^2 - \frac{\omega^2}{b}q +
    \frac{\omega^2}{2b^2}\right)\exp{(2bq)}-\frac{\omega^2}{2b^2}\right].
    \label{FiPoitSquar}
\end{equation}

One more example of the systems with `friction' is the nonlinear
Hirota oscillator playing an extremely important role in the study
of nonlinear evolution equations and dynamical
systems~\cite{Hirota,Laptev}. Its equation of motion and the
Lagrange function are
\begin{equation}
    \ddot{\phi}+(1+{\dot{\phi}}^2)\tan{\phi}=0
    \label{HirotaEq},
\end{equation}
\begin{equation}
    L=\dot{\phi}\arctan{\dot{\phi}}-\frac{1}{2}\ln(1+{\dot{\phi}}^2)+\frac{1}{2}
    \ln{\cos{\phi}}^2.
    \label{HirotaLagr}
\end{equation}

So, we have too many examples of equations containing the time
derivatives of generalized coordinates and we should look for some
explanation for this phenomenon. As a matter of fact this
``strange'' situation have been explained more than a century ago.

The mathematical approach \cite{Morandi} formulates the inverse
problem of the calculus of variations as the problem of finding
conditions, ensuring that a given system of differential equations
of motion coincides with the system of Euler-Lagrange equations of
an integral variational functional. This problem (in recent years also known
as the Sonin-Douglas problem \cite{Zenkov}) was first
considered by Sonin \cite{Sonin} for one second order ordinary differential
equation in  1886 (the almost forgotten paper). He proved
that every second order equation has a Lagrangian. Then the same
idea and approach appeared  in 1894 \cite{Darboux}. Later, it was
shown \cite{Morandi,Currie} that for one-dimensional systems in
context of the inverse problem of Lagrangian dynamics and
non-uniqueness of Lagrangian there are infinitely many Lagrangians
which result in the same trajectory in the configuration space for
any second-order differential equation. They are not, of course,
canonically equivalent in the usual sense, since they may very well
give different second order equations for $p$ and thus different
orbits in the phase space.

Different Lagrangian descriptions of the same system engender
different `energies',
$E  = \dot{q}\partial L/\partial \dot{q}
- L$. It turns out that,
depending on a particular choice of coordinates for its
Lagrangian description, a given dynamical system may be regarded
either as dissipative or not.

For $n\geqslant 2$,
the Lagrangian description is generically
unique \cite{Morandi,Douglas}, if there is any.

More exactly the Sonin result is as follow. Any second-order
differential equation,
\begin{equation}
    \ddot{q}-F(q,\dot{q},t)=0,
\end{equation}
can be presented in the Lagrangian form with the Jacobi last
multiplier $\mathcal{M }$
\begin{equation}
    \mathcal{M }\,(\ddot{q}-F(q,\dot{q},t))=0\quad\Leftrightarrow \quad \frac{d}{dt}
    \frac{\partial L}{\partial \dot{q}} - \frac{\partial L}{\partial
   q} =  0.
\end{equation}
The  multiplier $\mathcal{M }(q,\dot{q},t)$ satisfies the following
equation
\begin{equation}
      \frac{\partial \mathcal{M }}{\partial
   t}+\dot{q}\frac{\partial \mathcal{M }}{\partial
   q}+\frac{\partial \mathcal{M }F(q,\dot{q},t)}{\partial
   \dot{q}} =  0.
\end{equation}
After finding multiplier $\mathcal{M }$, the Lagrangian can be get
from the equation
\begin{equation}
      \frac{\partial^2 \mathcal{L }}{\partial
   \dot{q}^2}=\mathcal{M }.
\end{equation}

It is easy to verify that equations (\ref{EqDampedHO}) (with
constant parameters), (\ref{FiPoitSquarEq}) and (\ref{HirotaLagr})
have the following multipliers $M\exp(2\lambda t),\,m \exp(-2bq)$ and
$1/(1+\dot{\varphi}^2)$ correspondingly.

So, in this way, from the mathematical viewpoint of the  Lagrangian
and Hamiltonian approaches, there is nothing paradoxical in the fact
that the same Hamiltonian system, the generalized harmonic
oscillator in our case, is canonically equivalent to the two
different systems: the usual plane mathematical pendulum and the
damped harmonic oscillator. Nevertheless, in some physical
scientific circles there is still a popular belief that the
Lagrangian and Hamiltonian approaches are not suitable for the
consideration of dissipative systems.
Unfortunately, this  point of view is deeply rooted
and many authors keep on inventing
new formulations of the principle of the least action
(e.g., \cite{Tarasov08,Galley}).

\section{Conclusion\label{Concl}}

We studied the motion of planar linearized mathematical pendulum
with slowly varying parameters (the mass, the length of the
suspension string and the speed of the suspension point). The
Hamiltonian of the pendulum was cast into the form of the
Hamiltonian for the GHO. Thus, we give
the example of the simple Hamiltonian system
described by the equations of the GHO. The paradoxical feature of
the result is that the same Hamiltonian system, the GHO in this
case, can be canonically equivalent to the two different systems, to
the planar mathematical pendulum (the Hamiltonian system) and to the
damped harmonic oscillator (the dissipative system with time
dependent Lagrangian). This observation disputes the separation of
dynamical systems into two classes of the dissipative and
Hamiltonian ones. In our opinion the dividing line might be, for
example, between the systems which are invariant and non-invariant
with respect to the time reversal transformation.




\appendix

\section{}

Another convenient form of Hamiltonian (\ref{Hamiltonian1}) is its
complex form ${\cal H }^z$. It can be obtained after changing the
variables $Q$ and $P$ by $z$ and $\overline z$.
\begin{equation}
    Q= \sqrt{{\gamma\over {2 \omega}} }(z+ \overline z),\ \ \ \ \
     P={1\over \sqrt {2 \omega \gamma}} \big[(i \omega -\beta)z -
(i \omega +\beta)\overline z \big] \ . \label{Q Compl}
\end{equation}
The generating function $ F(Q,z)$ of the transformation $(P,Q)\to
(-i\overline z, z)$ is get by integration of the equations
$\displaystyle{ { {\partial  F } \over {\partial Q}}=P(Q,z)} $ and
${ \displaystyle{{\partial  F } \over {\partial z}} =i \overline
z}(Q,z)$ deduced from the differential identity characterizing a
canonical transformation
\begin{equation}
PdQ -H dt=-i {\overline z} dz -{\cal H }^z dt +dF\  .
\end{equation}
Taking into account the relations (\ref{Q Compl}) one gets
\begin{equation}
\label{F}
 F(Q,z) = - {{i \omega + \beta} \over {2\gamma}} Q^2 + i \sqrt{\frac{2 \omega}{\gamma} }
Qz - {1\over 2} i z^2 \  .
\end{equation}
The Hamiltonian ${\cal H }^z$ for the new conjugate variables
$(-i\overline z,z)$ is then obtained from the relation
$\displaystyle{ {\cal H }^z = H  + {\partial F \over
\partial t}}$ .
Its expression reads
\begin{equation}
\label{Hz}
 {\cal H }^z = \omega  z \overline z +  { {i\dot
\omega } \over {4 \omega }} ( z ^2 - \overline z ^2) -
 { {\dot \beta} \over {4 \omega} }(z +  \overline z )^2+
 {{\dot \gamma} \over {4 \omega \gamma} }[-i\omega(z^2 -\overline
 z^2 )+\beta(z +\overline z )^2].
\end{equation}
One can verify that the Hamilton equations
\begin{equation}
\dot z = { {\partial {\cal H }^z } \over {\partial (-i\overline
z)}}\ \  ,\ \ -i\dot {\overline z} = -{ {\partial {\cal H }^z  }
\over {\partial  z}}\
\end{equation}
indeed give correct equations for $z$ and the corresponding one for
$\overline z$ after substitution (\ref{Q Compl}) in (\ref{HamEqs}).
Conservation of the Poisson brackets
$\{Q,P\}_{QP}=\{Q,P\}_{z,-i\overline z}=1$ of the transformations
(\ref{Q Compl}) manifests about their canonicity. 

The multiplier $\omega   -
 { {\dot \beta} / {2 \omega} }+
 {{\dot \gamma \beta} / {2 \omega \gamma} }$ of the term $z \overline
 z$  in Eq. (\ref{Hz}) gives once again the result (\ref{D&GPhases}).


\section{ }
The other way to obtain the adiabatic invariant and phases is as
follow. After presenting the term $mlv\phi\dot{\phi}$ in the Lagrangian
(\ref{Lagrangian}) in the form $(1/2)d(mlv\phi^2)/dt -
(\phi^2/2)d(mlv)/dt$ and neglecting the total
derivative $(1/2)d(mlv\phi^2)/dt$, we get
\begin{equation}
    L=\frac{ml^2}{2}{\dot{\phi}}^2-mgl
        \left[1 -\frac{v}{g}
        \left(
        \frac{\dot{m}}{m}+
    \frac{\dot{v}}{v}
        \right)\right]
        \frac{{\phi}^2}{2},
\label{Lagrangian2}
\end{equation}
where we can immediately identify the squared frequency,
\begin{equation}
    \omega_{pend}^{2}=\frac{g}{l}\left[1 -\frac{v}{g}\left( \frac{\dot{m}}{m}+
    \frac{\dot{v}}{v}\right)\right].
\label{frequance}
\end{equation}
More rigorously, using Eqs. (\ref{Lagrangian2}) or
(\ref{Lagrangian}), we should write the Euler-Lagrange equation and
look for its solution in the form $\phi=r \cos \Theta$. Neglecting
quantities of the second order of $\epsilon$, we obtain simple
differential equations of the first order for $r$ and $\Theta$
containing first derivatives of the parameters. The solutions of
these equations yield  (\ref{Action}) and (\ref{D&GPhasesPend}).

\section*{References}


\begin{thebibliography}{20}

\bibitem{Wilczek} F.~Wilczek, A.~Shapere,
Geometric Phases in Physics,
World Scientific, 1989.

\bibitem{Vinitskii} S.I.~Vinitskii, V.L.~Derbov, V.M.~Dubovik, B.L.~Markovski,
Yu.P.~Stepanovskii,
Topological phases in quantum mechanics and polarization optics,
Sov. Phys. Usp. 33(6) (1990) 403-428.

\bibitem{Klyshko} D.N.~Klyshko,
Berry geometric phase in oscillatory processes,
Phys. Usp. 36(11) (1993) 1005-1019.

\bibitem{Malykin} G.B.~Malykin, S.A.~Kharlamov,
Topological phase in classical mechanics,
Phys. Usp. 46 (2003) 957-965.

\bibitem{Chruscinski} D.~Chruscinski, A.~Jamiolkowski,
Geometric Phases in Classical and Quantum Mechanics,
Springer Science and Business Media, 2004.

\bibitem{Berry84} M.V.~Berry,
Quantal Phase Factors Accompanying Adiabatic Changes,
Proc. Roy. Soc. A392 (1984) 45-57.

\bibitem{Hannay} J.H.~Hannay,
Angle variable holonomy in adiabatic excursion of an integrable Hamiltonian,
J. Phys. A: Math. Gen. 18 (1985) 221-230.

\bibitem{Berry85} M.V.~Berry,
Classical adiabatic angles and quantal adiabatic phase,
J. Phys. A: Math. Gen. 18 (1985) 15-27.

\bibitem{Dykhne60} A.M.~Dykhne,
Quantum Transitions in the Adiabatic Approximation,
Soviet Physics JETP 11 (1960) 411-415.

\bibitem{Fedoruk} M.V.~Fedoruk,
Asymptotic Analysis: Linear Ordinary Differential Equations,
translated from the Russian by Andrew Rodick,
Springer, Berlin, New York, 1993.

\bibitem{Bliokh} K.Yu.~Bliokh,
The appearance of a geometric-type instability
in dynamic systems with adiabatically varying parameters,
J. Phys. A: Math. Gen. 32 (1999) 2551-2565.

\bibitem{Usatenko}O.V.~Usatenko, G.P.~Provost, G.~Vallee,
A Comparative study of the Hannay's angles associated with a damped
harmonic oscillator and a generalized harmonic oscillator,
J. Phys. A: Math. Gen. 29 (1996) 2607-2610.

\bibitem{Razavy} M.~Razavy,
Classical and Quantum Dissipative Systems,
Imperial College Press, 2006.

\bibitem{Smith} C.E.~Smith,
Lagrangians and Hamiltonians with friction,
Journal of Physics: Conference Series 237 (2010) 012021.

\bibitem{Kobe} D.H.~Kobe, J.~Zhu,
Generalized Hanney angle for the most general
time-dependent harmonic oscillator,
Int. J. Mod. Phys. B 07 (1993) 4827-4840.


\bibitem{Slutskin64} A.A.~Slutskin,
Motion of a one-dimensional nonlinear
oscillator under adiabatic conditions,
Soviet Physics JETP 18 (1964) 676.

\bibitem{Caldirola} P.~Caldirola,
Forze non conservative nella meccanica quantistica,
Nuovo Cimento 18 (1941) 393-400.

\bibitem{Kanai} E.~Kanai,
On the quantization of dissipative systems, Prog. Theor. Phys. 3
(1948) 440-442.

\bibitem{Landau} L.D.~Landau, E.M.~Lifshits,
Mechanics, Butterworth-Heinemann, 1976.

\bibitem{Lanczos} C.~Lanczos,
Variational principles of mechanics,
Dover publications, Inc. New York, 1986.

\bibitem{Tarasov94} V.E.~Tarasov,
Quantum dissipative systems I. Canonical quantization and quantum
Liouville equation, Theoretical and mathematical physics 100 (1994)
402-417.

\bibitem{Tarasov08} V.E.~Tarasov,
Quantum mechanics of non-Hamiltonian and dissipative systems,
Elsevier Science, 2008.

\bibitem{Goldstein} H.~Goldstein, Classical mechanics, 3rd ed.,
Pearson Education Lim., 2014.

\bibitem{Hirota} R.~Hirota,
Exact N-Soliton Solution of Nonlinear Lumped Self-Dual Network Equations,
J. Phys. Soc. Jpn. 35 (1973) 289-294.

\bibitem{Laptev} D.V.~Laptev, Classical Energy Spectrum of the Hirota Nonlinear Oscillator,
J. Phys. Soc. Jpn. 82 (2013) 044005.

\bibitem{Morandi} G.~Morandi, C.~Ferrario, G.Lo.~Vecchio, G.~Marmo, C.~Rubano,
The inverse problem in the calculus of variations and the geometry
of the tangent bundle, Physics Reports 188 (1990) 147-284.

\bibitem{Zenkov} Zenkov, D. (ed.): The Inverse Problem of the Calculus of
Variations, Local and Global Theory. Atlantis Press, Amsterdam
(2015).

\bibitem{Sonin}  N. J. Sonin, About determining maximal and minimal properties of plane
curves, Warsawskye Universitetskye Izvestiya (1886) (1-2), 1-68 (in
Russian); English translation by R. Matsyuk, Lepage Inst. Archive,
No. 1 (2012), 1-42.

\bibitem{Darboux} Darboux, G.: Lecons sur la theorie generale des surfaces.
Gauthier-Villars, Paris (1894)


\bibitem{Currie} D.G.~Currie and E.J.~Saletan,
q-Equivalent Particle Hamiltonians. I. The Classical One-dimensional Case,
J. Math. Phys. 7 (1966) 967-974.

\bibitem{Douglas} Douglas, J.: Solution of the inverse problem of the calculus of
variations. Trans. AMS 50, 71-128 (1941).


%

\bibitem{Galley} C.G.~Galley,
Classical Mechanics of Nonconservative Systems, Phys. Rev. Lett. 110
(2013) 174301.



\end{thebibliography}
\end{document}